\documentclass[12pt]{article}
\usepackage{amssymb}
\usepackage[onehalfspacing]{setspace}

\newtheorem{theorem}{Theorem}

\newtheorem{proposition}[theorem]{Proposition}

\input{tcilatex}
\begin{document}

\title{Anabolic Persuasion\thanks{%
We acknowledge financial support from the Foerder Institute. We thank Martin
Cripps, Michael Crystal, Israel Halperin, Nathan Hancart, Ron Peled, Mickey
Scheinowitz and seminar audiences at UCL and Caltech for helpful comments.
We owe special thanks to Oliver Fishman, who produced a fitness video for
this paper (https://youtu.be/onUrRs3kHIE).}}
\author{Kfir Eliaz and Ran Spiegler\thanks{%
Eliaz: School of Economics, Tel-Aviv University and David Eccles School of
Business, University of Utah. E-mail: kfire@tauex.tau.ac.il. Spiegler:
School of Economics, Tel-Aviv University and Economics Dept., University
College London. E-mail: rani@tauex.tau.ac.il.}}
\maketitle

\begin{abstract}
We present a model of optimal training of a rational, sluggish agent. A
trainer commits to a discrete-time, finite-state Markov process that governs
the evolution of training intensity. Subsequently, the agent monitors the
state and adjusts his capacity at every period. Adjustments are incremental:
the agent's capacity can only change by one unit at a time. The trainer's
objective is to maximize the agent's capacity - evaluated according to its
lowest value under the invariant distribution - subject to an upper bound on
average training intensity. We characterize the trainer's optimal policy,
and show how stochastic, time-varying training intensity can dramatically
increase the long-run capacity of a rational, sluggish agent. We relate our
theoretical findings to \textquotedblleft periodization\textquotedblright\
training techniques in exercise physiology.\bigskip \bigskip
\end{abstract}

\section{Introduction}

Economists have a long tradition of invading other academic disciplines.
Lazear (2000) celebrates this so-called economic imperialism, demonstrating
its value for such diverse fields as sociology, criminology and
organizational behavior. Proponents of economic imperialism maintain that
the ideas of individual rationality, forward-looking behavior and
equilibrium help us understand empirical regularities and guide policy
interventions. As Becker (1976, p. 8) wrote: \textquotedblleft I have come
to the position that the economic approach is a comprehensive one that is
applicable to all human behavior\textquotedblright . Recently, economists
applied the imperialistic approach to the field of epidemiology, in the
context of the Covid-19 pandemic (Acemoglu et al. (2020)).

This paper carries the imperialistic approach to a new territory. It applies
tools from economic theory to the field of \textit{exercise physiology},
which studies the body's response and adaptation to exercise to maximize
human physical potential (for an introduction to this field, see Glass et
al. (2014)). Specifically, we focus on the question of how the body's muscle
mass responds to patterns of physical exercise. Muscle mass adapts to
physical stimuli, and the economic approach seeks to describe this
adaptation as the result of maximizing behavior. We demonstrate that by
modeling the body as a forward-looking optimizing agent, we gain insights
into the effectiveness of popular physical training strategies.

To describe the body as an optimizing agent, we need to specify its
preferences. On the one hand, maintaining muscle mass is costly in terms of
energy expenditure (Zurlo et al. (1990)). On the other hand, if muscle mass
is too low relative to the demands of exercise, the body may incur the
energy costs of repairing torn muscle tissue or inflammation (see
Frankenfield (2006) and Faulkner et al. (1993)). Moreover, if the body lacks
adequate muscle mass, it will not be able to complete the required physical
task. It is plausible to assume that the body will internalize this
performance gap as a cost. This cost can be interpreted in terms of
psychological motivation, which itself may originate from evolutionary
survival pressures (see Sagar and Stoeber (2009) and Lieberman (2015)). A
more motivated trainee will record the performance gap as a larger cost
relative to the muscle maintenance cost.

In a dynamic environment where the intensity of exercise changes
stochastically over time, a key ingredient in modeling the body as an
optimizing agent is its expectation of future demands. Here, too, we follow
the economist's standard recipe and assume that the body has rational
expectations - i.e., it knows the stochastic process that governs the future
evolution of physical exercise, possibly as a result of some previous
adaptive-learning phase.

Using these ingredients, we construct the following stylized discrete-time
model. A \textquotedblleft trainer\textquotedblright\ commits ex-ante to a
strategy, which is a stochastic process that governs the evolution of
exercise intensity. We restrict ourselves to stochastic processes that
follow a finite-state Markov chain. Average intensity (according to the
chain's invariant distribution) cannot exceed some integer $\mu $ more than
negligibly. The parameter $\mu $ represents a \textquotedblleft budget
constraint\textquotedblright\ that limits the amount of resources that can
be devoted to physical training.

Following the trainer's choice of strategy, at every subsequent period, the
body (referred to as an \textquotedblleft agent\textquotedblright ) monitors
the state of the trainer's Markov process and chooses its muscle mass. We
assume that the body can only make \textit{incremental adjustments} to its
muscle mass: at every period it can only change the mass by $-1$, $0$ or $1$
units. The body is an expected discounted utility maximizer, with a periodic
payoff function that trades off the maintenance cost of muscle mass and the
excess intensity of current physical exercise relative to current mass. We
measure muscle mass and exercise intensity with the same units, such that
excess intensity (also referred to as the performance gap) is a simple
difference between the two numbers. We impose the constraint that the agent
has a best-reply to the trainer's strategy that induces a Markov process
(over an extended state space that also includes muscle mass in the
definition of a state) with a $unique$ invariant distribution.

The trainer's problem is to choose the Markov process to maximize the
agent's long-run mass - evaluated according to its \textit{minimal}
realization under the invariant distribution. Our use of such a
\textquotedblleft max-min\textquotedblright\ criterion is justified by the
interpretation of muscle mass as a \textit{capability}: the higher the
agent's minimal long-run mass, the higher the intensity he can reliably
withstand in the long run.

The sluggish adjustment of muscle mass is a fundamental assumption in our
model. Exercise intensity can fluctuate wildly between periods, but clearly,
the body cannot change its muscle mass instantaneously to any level (see
DeFreitas et al. (2011) and Counts et al. (2017)). Our model is set up such
that if the body had perfect flexibility in adjusting its muscle mass, the
trainer's problem would be trivial: at every period, muscle mass would be
set such that excess intensity would be exactly zero, and the long-run
average mass will be at most $\mu $. This is also the minimal long-run mass
that the trainer can attain with a constant-intensity policy. Under such a
policy, the distinction between sluggish and flexible agents is irrelevant.
The question is whether using some non-degenerate Markov process will enable
the trainer to outperform this benchmark when muscle adjustment is sluggish.

The trainer's problem is similar in spirit to Bayesian persuasion (Kamenica
and Gentzkow (2011)). In a persuasion problem, the sender wants to increase
the receiver's action; in our model the trainer wants to increase the
agent's muscle mass. In a persuasion problem, the sender commits to a signal
function; in our problem, the trainer commits to a Markov process. In a
persuasion problem, the receiver's response to a signal realization is
dictated by Bayesian updating; in our model, the agent's response to a
realized state is constrained by sluggish adjustment. Finally, in a
persuasion problem, the sender's ability to attain his objective is
constrained by Bayes plausibility, which requires the average posterior
belief to equal the prior; in our model, the trainer is constrained by the
average intensity limit $\mu $.

Our analysis focuses on two extreme cases in terms of the agent's discount
factor. We begin by analyzing a myopic agent, whose adjustment rule is
mechanical and independent of the trainer's strategy:\ mass moves up (down)
a notch when current intensity is above (below) current mass. In this case,
the trainer cannot attain a minimal long-run mass above $2\mu -1.$ He can
implement this upper bound using a two-state Markov process with intensity
levels $0$ and $2\mu $; the transition from $0$ to $2\mu $ is deterministic,
while the transition from $2\mu $ to $0$ occurs with $near$ certainty. This
random element ensures that regardless of the initial muscle mass, the agent
eventually oscillates between mass levels $2\mu $ and $2\mu -1.$

We next turn to an arbitrarily patient agent. In this case, the trainer
cannot attain a minimal long-run muscle mass above $\mu /c-1$, where $c$ is
the maintenance cost per unit of muscle mass (we assume that $\mu /c$ is an
integer, for convenience). The trainer can implement this bound using a
two-state Markov process with intensity levels $0$ and $\mu /c$; one of the
transitions between the two states occurs with certainty (which one depends
on the value of $c$). This policy ensures that regardless of the initial
mass, the agent eventually oscillates between muscle mass levels $\mu /c$
and $\mu /c-1$. The transition probabilities are calibrated to make the body
nearly indifferent to lowering muscle mass below $\mu /c-1$.

Thus, when the agent is sluggish, a properly designed stochastic training
program can increase long-run mass substantially relative to the
flexible-adjustment benchmark. Our results suggest a rationale for the
popular training technique of \textit{periodization}, which structures the
training regimen as a cycle with phases of high intensity physical load and
recovery phases of low intensity. Since it first began in the 1960s, this
methodology has gained popularity and is currently the dominant technique
used by professional athletes. Numerous studies have documented the success
of periodization in terms of increased muscle mass, increased muscle
strength, greater endurance and athletic performance (Bompa and Buzzichelli
(2018)).\footnote{%
For recent discussions of various periodization techniques, see Issurin
(2010), Kiely (2012) and Kiely\ et al. (2019).} While the physiological
literature offers biological explanations for the superiority of a cyclical
training design (e.g., see Issurin (2019)), our results provide a
complementary perspective: the effectiveness of periodization techniques may
stem from rational yet sluggish adaptation to fluctuations in physical
stimuli.

Although our paper strictly adheres to the model's exercise-physiology
interpretation, its abstraction enables other interpretations. For example, $%
m$ and $d$ may represent cognitive capacity and the intensity of cognitive
activity, such that our results can be viewed in terms of programs for
maintaining cognitive skills. Moving to more conventionally economic
settings, we can view the agent as an \textit{organization} like a military
or an emergency-management agency. The mission of such organizations is to
maintain a level of preparedness against unexpected challenges. This can be
achieved with a suitably designed regimen of \textit{drills}. Sluggish
adjustment is a natural assumption in this setting: organizations cannot
drastically improve their level of preparedness overnight; and likewise,
deteriorating preparedness tends to be gradual. Our analysis sheds light on
the optimal design of a drill program for such organizations. More
generally, we find the optimal design of a stochastic process for a sluggish
agent to be an interesting (and, to our knowledge, new) problem from an
abstract economic-theory perspective.

\section{The Model}

We consider a principal-agent model, in which the principal is referred to
as a\textit{\ \textquotedblleft }trainer\textquotedblright . We interpret
the agent as a physiological system that is trained to increase its
capacity. The trainer commits ex-ante to a pair $(P,f)$, where $P$ is a
discrete-time, Markov process over some finite set of states $S$, and $%
f:S\rightarrow 
\mathbb{N}
_{+}$ is an output function that assigns a challenge level to every state $%
s\in S$. We denote by $d_{t}$ the challenge level in period $t$. When there
is no risk of confusion we will replace the notation $f(s)$ with $d(s).$

We impose the following constraints on $(P,f)$. First, $P$ has a unique
invariant distribution $\lambda _{P}$. Second,%
\[
\sum_{s\in S}\lambda _{P}(s)f(s)\leq \mu +\varepsilon 
\]%
where $\mu \geq 1$ is an integer and $\varepsilon >0$ is arbitrarily small.
That is, the long-run average challenge level cannot exceed $\mu $ by more
than a negligible amount (the approximate formulation of the constraint is
due to $\mu $ getting integer values).

The agent knows the trainer's choice of $(P,f).$ At every period $t,$ he
observes the realized state $s_{t}$ and then chooses a non-negative capacity
level $m_{t}\in \{m_{t-1}-1,m_{t-1},m_{t-1}+1\}$. Henceforth, we refer to $%
m_{t}$ as the agent's \textquotedblleft mass\textquotedblright\ at time $t$.
Let $m_{0}\in 
\mathbb{N}
_{+}$ be the agent's initial mass. The restricted choice set for $m_{t}$
reflects the sluggish adaptation of the agent's mass.

The agent is an expected discounted utility maximizer with discount factor $%
\delta $. His payoff at period $t$ is%
\[
-[cm_{t}+\max (0,d_{t}-m_{t})] 
\]%
where $c\in (0,1)$, $d_{t}=f(s_{t})$ and $s_{t}$ is the state of $P$ at
period $t$. The body's periodic cost incorporates two factors. First, $%
cm_{t} $ is the caloric maintenance cost of muscle mass $m_{t}$. Second, the
gap between $m_{t}$ and $d_{t}$ (when the latter is higher) represents a
performance shortfall because the agent's capacity is lower than the
challenge it faces.

Given $(P,f)$, the agent faces a Markov decision problem over an extended
state space, where the state at period $t$ is the pair $(s_{t},m_{t-1})$. We
impose the following additional constraint on the trainer: the extended
Markov process over $(s_{t},m_{t-1})$ that is induced by the agent's
best-reply to $(P,f)$ has a unique invariant distribution $\lambda
_{(P,f)}^{\ast }$. This ensures that the minimal and average long-run masses
are well-defined and independent of the initial condition $m_{0}$.

The trainer aims to maximize the agent's lowest muscle mass in the support
of the invariant distribution $\lambda _{(P,f)}^{\ast }$. The larger this
mass, the higher the challenge level that the agent is guaranteed to meet in
the long run. Formally, the trainer's problem can be stated as follows:%
\[
\max_{(P,f)}\min \{m\mid \lambda _{(P,f)}^{\ast }(s,m)>0\text{ for some }%
s\in S\} 
\]%
subject to the feasibility constraint%
\[
\sum_{(s,m)}\lambda _{(P,f)}^{\ast }(s,m)f(s)\lessapprox \mu 
\]%
The max-min criterion means that the trainer looks for the highest capacity
that the agent's body \textit{reliably} maintains in the long run. The
symbol $\lessapprox $ represents the requirement that average intensity
cannot exceed $\mu $ by more than a negligible amount.\medskip

\noindent \textit{Discussion of the model's interpretation}

\noindent The level of physical challenge $d$ can be interpreted in terms of
duration (e.g. the number of repetitions of a given exercise), load (e.g.
lifting weight) or effort (e.g. running speed).\footnote{%
See Steele (2014) and Steele et al. (2017) for discussions of these
different notions of intensity.} The stylized nature of our model abstracts
from such fine distinctions. However, the interpretation of the resource
constraint does depend on the meaning of $d$. If it represents exercise
duration, then $\mu $ is the average amount of time per period that the
trainee can devote to physical exercise. If, however, $d$ represents load or
effort, $\mu $ is perhaps better viewed as a parameter of the trainer's
problem than an exogenous resource constraint.

Our model endows the human body with rational expectations: it has knowledge
of $(P,f)$ when making its periodic decisions. The justification for this
assumption is that the body forms adaptive expectations based on a long
memory. We find it reasonable to assume that in the long run, the body will
learn finite-state Markov processes, especially when they have few states.

The adaptive-expectations rationale also underlies our restriction that the
trainer cannot condition $d_{t}$ on past realizations of $m.$ If he could,
he would have recourse to off-equilibrium threats. For instance, he could
incentivize the agent to increase muscle mass using a policy of zero on-path
challenges, sustained by a \textquotedblleft grim\textquotedblright\ threat
to switch to persistently extreme challenges if $m$ fails to go up. We find
such policies absurd in the physiological context and attribute this
absurdity to the implausibility of full-throttle rational expectations in
this content. We effectively rule out off-path threats by assuming that the
trainer does not condition on $m$. Under our alternative interpretation of
the agent as an organization, it is questionable whether the trainer can
monitor $m,$ which represents the organization's level of preparedness.%
\footnote{%
We conjecture that if the trainer can condition $d_{t}$ on $m_{t-1}$, the
results in our paper will not change.}\medskip

\noindent \textit{Benchmark: Completely flexible adjustment}

\noindent Suppose the agent could choose any $m_{t}\in 
\mathbb{N}
_{+}$ at every period. Then, since $c\in (0,1)$, he would choose $%
m_{t}=d_{t} $ at every $t$. This means that the long-run average of $m_{t}$
would coincide with the long-run average of $d_{t}$, which by assumption
cannot exceed $\mu $ more than negligibly. Therefore, the best the trainer
can do according to his max-min criterion is play a constant strategy $%
d_{t}=\mu $ at every period, such that the flexible agent's mass will be $%
\mu $ as well. The same deterministic process attains the same long-run mass
of $\mu $ also when the agent is sluggish (because the agent will eventually
reach this mass and stay there indefinitely). The question is whether the
trainer can outperform this benchmark with a non-degenerate Markov process.

\section{A Myopic Agent}

In this section we analyze the trainer's problem when $\delta =0$ - i.e.,
the agent is myopic.\medskip 

\begin{proposition}
\label{prop 1}Let $\delta =0$. Then:\medskip \newline
(i) For any trainer strategy, the minimal long-run mass induced by the
agent's best-reply is at most $2\mu -1$.\medskip \newline
(ii) This upper bound can be implemented by the following $(P,f)$. The
Markov process $P$ has two states, $H$ and $L$, and a transition matrix
given by%
\[
\begin{array}{ccc}
\Pr (s_{t}\rightarrow s_{t+1}) & L & H \\ 
L & 0 & 1 \\ 
H & \beta  & 1-\beta 
\end{array}%
\]%
where $\beta $ is arbitrarily close to $1$. The output function is $%
f(H)=2\mu $ and $f(L)=0$. In the $\beta \rightarrow 1$ limit, the invariant
mass distribution assigns probability $\frac{1}{2}$ to $m=2\mu $ and $m=2\mu
-1$.\medskip 
\end{proposition}

Thus, a slightly perturbed cyclic training program can dramatically increase
the minimal long-run mass of a myopic sluggish agent. When $\mu $ is large
(corresponding to a very sluggish agent, given that we normalized his
adjustment increment to $1$), the increase is by a factor of nearly $2$
relative to the flexible-agent benchmark.

The trainer's training regime\ approximately consists of alternating periods
of high intensity ($d=2\mu $) and rest ($d=0$). After a period of high
intensity training, there is a small chance $1-\beta $ that the
high-intensity episode will be repeated. This stochastic perturbation
ensures that the set of mass values $\{2\mu ,2\mu -1\}$ is absorbing: the
agent will reach it in finite time with probability one, regardless of $m_{0}
$.\pagebreak 

\noindent \textbf{Proof of part }$(i)$\textbf{\ of Proposition \ref{prop 1}}

\noindent The proof proceeds by a series of steps. Recall that we use the
notation $d(s)$ as a substitute for $f(s)$.\medskip

\noindent \textbf{Step 1: }\textit{The agent's strategy}

\noindent Consider the agent's move at period $t$, given the extended state $%
(s_{t},m_{t-1})$. A myopic agent will choose $m_{t}$ to minimize $%
cm_{t}+\max (0,d(s_{t})-m_{t})$. Therefore, we can immediately pin down the
agent's behavior, independently of the trainer's strategy. Since $c\in (0,1)$%
, we obtain the following: if $d(s_{t})>m_{t-1}$, the agent will choose $%
m_{t}=m_{t-1}+1$; if $d(s_{t})<m_{t-1}$, the agent will choose $%
m_{t}=m_{t-1}-1$; and if $d(s_{t})=m_{t-1}$, the agent will choose $%
m_{t}=m_{t-1}$. That is, the agent will always adjust his mass in the
direction of the current level of $d$. $\square $\medskip

Consider an arbitrary strategy for the trainer, which induces an extended
Markov process with a unique invariant distribution. Let $%
(m_{t-1},d_{t})_{t=1,2,...}$ be a possible sample path that results from the
extended process. By the unique-invariant-distribution requirement, the
extended process is ergodic. Therefore, the long-run frequency of every $%
(m,d)$ in the sample path coincides with the probability of this pair
according to the invariant distribution. Let $\lambda (m,d)$ denote the
probability of $(m,d)$ according to the invariant distribution, as well as
the frequency of $(m,d)$ in the sample path. Let $X$ be the set of recurrent
pairs $(m,d)$ in the sample path. Partition $X$ into three classes:%
\begin{eqnarray*}
X^{+} &=&\{(m,d)\in X\mid d>m\} \\
X^{-} &=&\{(m,d)\in X\mid d<m\} \\
X^{0} &=&\{(m,d)\in X\mid d=m\}
\end{eqnarray*}%
\pagebreak 

\noindent \textbf{Step 2: }$\lambda $\textit{\ satisfies}%
\begin{equation}
\sum_{(m,d)\in X^{+}}\lambda (m,d)(m+1)=\sum_{(m,d)\in X^{-}}\lambda (m,d)m
\label{lambda}
\end{equation}%
\noindent Consider some period $t$ along the sample path such that $%
(m_{t},d_{t+1})\in X^{+}$. By definition, this pair is recurrent. Therefore, 
$m_{t}$ must be visited again in some later period. Let $t^{\prime }+1$ be
the earliest such period. Since $m$ moves only in one-unit increments, it
must be the case that $(m_{t^{\prime }},d_{t^{\prime }+1})\in X^{-}$ and $%
m_{t^{\prime }}=m_{t}+1$. We have thus defined a one-to-one mapping from
periods $t$ for which $(m_{t},d_{t+1})\in X^{+}$ to periods $t^{\prime }$
for which $(m_{t^{\prime }},d_{t^{\prime }+1})\in X^{-}$, such that $%
m_{t^{\prime }}=m_{t}+1$. In a similar way, we can define a one-to-one
mapping from periods $t$ for which $(m_{t},d_{t+1})\in X^{-}$ to periods $%
t^{\prime }$ for which $(m_{t^{\prime }},d_{t^{\prime }+1})\in X^{+}$, such
that $m_{t^{\prime }}=m_{t}-1$. It follows that%
\[
\lim_{T\rightarrow \infty }\frac{\sum_{t=1}^{T}\mathbf{1}[(m_{t},d_{t+1})\in
X^{+}]\cdot (m_{t}+1)}{T}=\lim_{T\rightarrow \infty }\frac{\sum_{t=1}^{T}%
\mathbf{1}[(m_{t},d_{t+1})\in X^{-}]\cdot m_{t}}{T}
\]%
which can be rewritten as (\ref{lambda}). $\square $\medskip 

\noindent \textbf{Step 3: }\textit{The average long-run }$m$\textit{\ is at
most }$2\mu $\textit{\ (approximately)}

\noindent The long-run average of $m$ induced by the trainer's strategy can
be written as%
\begin{equation}
\mathbb{E}(m)=\sum_{(m,d)\in X^{+}}\lambda (m,d)m+\sum_{(m,d)\in
X^{-}}\lambda (m,d)m+\sum_{(m,d)\in X^{0}}\lambda (m,d)m  \label{Em}
\end{equation}%
By the feasibility constraint,%
\[
\sum_{(m,d)\in X^{+}}\lambda (m,d)d+\sum_{(m,d)\in X^{-}}\lambda
(m,d)d+\sum_{(m,d)\in X^{0}}\lambda (m,d)d\lessapprox \mu 
\]%
By definition, $d\geq m+1$ for every $(m,d)\in X^{+}$, $d\geq 0$ for every $%
(m,d)\in X^{-}$, and $d=m$ for every $(m,d)\in X^{0}$. Therefore,%
\[
\sum_{(m,d)\in X^{+}}\lambda (m,d)(m+1)+\sum_{(m,d)\in X^{-}}\lambda
(m,d)\cdot 0+\sum_{(m,d)\in X^{0}}\lambda (m,d)m\lessapprox \mu 
\]%
This means that%
\[
\sum_{(m,d)\in X^{+}}\lambda (m,d)m\leq \sum_{(m,d)\in X^{+}}\lambda
(m,d)(m+1)\lessapprox \mu -\sum_{(m,d)\in X^{0}}\lambda (m,d)m 
\]%
By (\ref{lambda}), it follows that%
\[
\sum_{(m,d)\in X^{-}}\lambda (m,d)m\lessapprox \mu -\sum_{(m,d)\in
X^{0}}\lambda (m,d)m 
\]%
as well. Plugging the last two inequalities in (\ref{Em}), we obtain%
\[
E(m)\lessapprox 2\mu -\sum_{(m,d)\in X^{0}}\lambda (m,d)m\leq 2\mu 
\]%
$\square $\medskip

\noindent \textbf{Step 4: }\textit{The minimal long-run }$m$\textit{\ is at
most }$2\mu -1$

\noindent Suppose the long-run distribution over $d$ is degenerate at some $%
d^{\ast }$. Therefore, $d^{\ast }\lessapprox \mu $. The agent's myopic
best-reply implies that eventually, his mass coincides with $d^{\ast }$. It
follows that to reach a minimal long-run mass above $\mu $, the long-run
distribution over $d$ must assign positive probability to at least two
values. This means there are infinitely many periods $t$ in which $d_{t}\neq
m_{t-1}$. By myopic best-replying, this precludes the possibility that the
long-run distribution over $m$ is degenerate. Since the long-run average of $%
m$ cannot exceed $2\mu $ by more than an infinitesimal amount, there must be
infinitely many periods $t$ in which $m_{t}\leq 2\mu -1$. This completes the
proof of part $(i)$. $\square $\medskip 

\noindent \textbf{Proof of part }$(ii)$\textbf{\ of Proposition \ref{prop 1}}

\noindent Consider the trainer's strategy described in part $(ii)$ of the
statement of the result. As long as $\beta \in (0,1)$, the Markov process
over $m$ that is induced by the strategy and the agent's best-reply (given
by Step 1) has a unique invariant distribution, with $m=2\mu $ and $m=2\mu -1
$ being the only recurrent mass values. The reason is that if $m_{t}>2\mu $, 
$m_{t+1}=m_{t}-1$ with certainty; if $m_{t}<2\mu -1$, there is a positive
probability that there will be a streak of realizations $d=2\mu $ such that $%
m$ will keep adjusting upward until it reaches $m=2\mu $; and finally, if $%
d_{t}=0$ then $d_{t+1}=2\mu $ for sure, which means that once $m$ hits $2\mu 
$ and later goes down to $2\mu -1$, it will return to $2\mu $ immediately in
the next period. As the exogenous upper bound on average intensity gets
arbitrarily close to $\mu $, $\beta $ can be made arbitrarily close to one.
In the $\beta \rightarrow 1$ limit, the invariant distribution over $m$
assigns probability $\frac{1}{2}$ to each of the values $m=2\mu $ and $%
m=2\mu -1$. $\blacksquare $

\section{A Patient Agent}

In this section we characterize the solution to the trainer's problem when
the agent is forward-looking and arbitrarily patient. For expositional
convenience, we assume $\mu /c$ is an integer.\medskip 

\begin{proposition}
\label{prop2}Let $\delta $ be arbitrarily close to $1$. Then:\medskip 
\newline
(i) The minimal long-run mass at the solution to the trainer's problem is at
most $\mu /c-1$.\medskip \newline
(ii) This upper bound can be implemented by $(P,f)$ with the following
properties. The Markov process $P$ has two states, $H$ and $L$, and a
transition matrix given by%
\[
\begin{array}{ccc}
\Pr (s_{t}\rightarrow s_{t+1}) & L & H \\ 
L & 1-\alpha  & \alpha  \\ 
H & \beta  & 1-\beta 
\end{array}%
\]%
where $\alpha =1$ if $c\geq \frac{1}{2}$, $\beta =1$ if $c<\frac{1}{2}$, and 
$\alpha /(\alpha +\beta )$ is arbitrarily close to $c$ from above. The
output function is $f(H)=\mu /c$ and $f(L)=0$. In the $\alpha /(\alpha
+\beta )\rightarrow c$ limit, the invariant mass distribution assigns
probability $c$ to $m=\mu /c$ and probability $1-c$ to $m=\mu /c-1$.\medskip 
\end{proposition}

The upper bound on the agent's minimal long-run mass is higher than in the
myopic benchmark whenever $c<\frac{1}{2}$. Moreover, it gets arbitrarily
high when $c\rightarrow 0$. As $c$ gets closer to one, the highest minimal
long-run mass approaches the flexible-agent benchmark $\mu $.\footnote{%
Because $\mu /c$ is an integer, we rule out the possibility that $c$ is
arbitrarily close to one. In that case, the trainer cannot outperform the
flexible-agent benchmark of $\mu $.}

The Markov process that attains the upper bound is similar to the one in
Section 3. The main difference is that persistence of one of the two states
occurs with non-vanishing probability. When $c<\frac{1}{2}$, a
\textquotedblleft rest period\textquotedblright\ (corresponding to the state 
$L$) is followed by another one with probability approximately equal to $%
(1-2c)/(1-c)$. When $c>\frac{1}{2}$, a high-intensity period (corresponding
to the state $H$) is followed by another one with probability $(2c-1)/c$.

Compare this with Section 3. The myopic agent only responds to current
realizations of $d$. In contrast, the patient agent reacts to the trainer's
entire continuation strategy. When $c<\frac{1}{2}$, the trainer's program
allows for a streak of $d=0$ realizations. When this happens, the agent does
not lower his mass below $\mu /c-1$ because he takes into account the future
loss $d-m$ in the event that $d$ switches to $d=\mu /c$. The trainer designs
the transition probabilities such that the patient agent's intertemporal
trade-offs lead him to be nearly indifferent between lowering his mass and
remaining at $m=\mu /c-1$. In contrast, the myopic agent cannot be made
indifferent when faced with a streak of $d=0$ realizations: he repeatedly
lowers his mass. This difference enables the trainer to achieve a higher
minimal long-run mass when the agent is patient, as long as $c<\frac{1}{2}$.

We now turn to the proof of Proposition \ref{prop2}. In our proof of part $%
(i)$, we actually prove a somewhat stronger result: in order to attain a
strictly positive minimal long-run mass, the $average$ long-run mass cannot
exceed $\mu /c-1+c$. The Markov process we construct in part $(ii)$
approximates this upper bound. This means that among all trainer strategies
that attain the minimal long-run mass of $\mu /c-1$, this process cannot be
outperformed in terms of average mass.\medskip 

\noindent \textbf{Proof of part }$(i)$\textbf{\ of Proposition \ref{prop2}}

\noindent Let $p$ be the unique invariant distribution over $(d_{t},m_{t})$
that results from the trainer's strategy and the agent's best-replying
strategy. (Note the different time subscripts of $d$ and $m$, compared with
the proof of Proposition \ref{prop 1} in Section 3; our different notation
highlights this difference.) We abuse notation and write $p(d)$, $p(m)$ and $%
p(d\mid m)$ to represent marginal and conditional distributions induced by $p
$. As in the myopic-agent case, we first derive an upper bound on the
expected mass according to $p$, which we use to derive the upper bound on
the minimal long-run mass. Then, we show how to implement this upper bound.

In Section 2, we saw that the trainer can implement a minimal long-run mass
of at least $\mu $ (by playing $d=\mu $ at every period). Therefore, we take
it for granted that the minimal value of $m$ in the support of $p$ is at
least $\mu \geq 1$.\medskip

\noindent \textbf{Step 1: }$p(d>0)\geq c$

\noindent Consider the following deviation by the agent. Pick some period-$t$
history for which $m_{t-1}\geq 1$ is at the lowest value according to $p$.
Therefore, $m_{t}=m\in \{m_{t-1},m_{t-1}+1\}$. At this history, the agent
deviates to $m_{t}^{\prime }=m-1$. Subsequently, the agent behaves according
to his original strategy \textit{as if the deviation did not occur}.

This deviating strategy induces an invariant distribution $p^{\prime }$ such
that for every $(d,m)$ in the support of $p$, $p^{\prime }(d,m-1)=p(d,m)$.
Therefore, the deviation saves $c$ at every period, but raises costs by one
unit per period whenever $d\geq m$ under the original strategy. In order for
this deviation to be unprofitable for an arbitrarily patient agent, it must
be the case that $p(d\geq m)\geq c$. Since $m>0$ with probability one, $%
p(d>0)\geq p(d\geq m)$, hence $p(d>0)\geq c$\textbf{. }$\square $\medskip

\noindent \textbf{Step 2: }\textit{The expectation of }$m$\textit{\
according to }$p$\textit{\ is at most }$\mu /c-1+c$

\noindent Assume the contrary. Then, the agent's average long-run cost
exceeds 
\[
c\cdot \lbrack \frac{\mu }{c}-1+c]=\mu -c(1-c) 
\]%
Now consider a deviation to the following strategy. Descend from $m_{0}$ to $%
m=0$, and then implement the following rule: $m_{t}=0$ whenever $d_{t}=0$,
and $m_{t}=1$ whenever $d_{t}>0$. When the agent is arbitrarily patient, the
average long-run cost from this strategy is approximately%
\begin{eqnarray*}
p(d &=&0)\cdot 0+p(d>0)\cdot \lbrack c+\sum_{d>0}p(d\mid d>0)d-1] \\
&\lessapprox &p(d>0)(c-1)+\mu
\end{eqnarray*}%
Since $c<1,$ Step 1 implies that%
\[
p(d>0)(c-1)+\mu <\mu -c(1-c) 
\]%
such that the deviation is profitable, a contradiction. $\square $\medskip

\noindent \textbf{Step 3: }\textit{The minimal long-run mass is at most }$%
\mu /c-1$

\noindent Since $\mu /c$ is an integer, $\mu /c-1+c$ is not an integer.
Hence, in order for the average long-run cost to be weakly below\ $\mu /c-1+c
$, the minimal long-run mass cannot exceed $\mu /c-1$.\footnote{%
The proof of this step utilizes the convenient assumption that $\mu /c$ is
an integer. An alternative proof that does not rely on this assumption is
analogous to Step 4 in the proof of Proposition \ref{prop 1}.} $\square $%
\medskip 

\noindent \textbf{Proof of part }$(ii)$\textbf{\ of Proposition \ref{prop2}}

\noindent Consider the strategy described in the statement of part $(ii)$.
Our objective is to show that given this strategy, there is a best-reply for
the agent such that for every sufficiently high $t$, $m_{t}=\mu /c$ whenever 
$s_{t}=H$ and $m_{t}=\mu /c-1$ whenever $s_{t}=L$.

Since the agent faces a Markovian decision problem with an extended state
space $(s,m)$, there exists a best-reply that is Markovian with respect to
this state space. To derive such a best reply, we proceed in four
steps.\medskip

\noindent \textbf{Step 1:} \textit{There is no best-reply in which the
invariant distribution assigns probability one to a single} $m$.

\noindent \textit{Proof.} Assume the contrary. If $m<\mu /c$, then it is
profitable for the agent to deviate to a strategy that plays $m+1$ whenever $%
s=H$ and $m$ whenever $s=L$. Likewise, if $m>0$, it is profitable for the
agent to deviate to a strategy that plays $m$ whenever $s=H$ and $m-1$
whenever $s=L$. $\square $\medskip

\noindent \textbf{Step 2:} \textit{The set of recurrent values of }$m$%
\textit{\ (according to the unique invariant distribution induced by the two
parties' strategies) is a set of consecutive numbers }$\underline{m},%
\underline{m}+1,...,\overline{m}$\textit{, where }$\overline{m}\leq \mu /c$.

\noindent \textit{Proof}. The agent's sluggishness implies that if the agent
visits two non-adjacent masses $m$ and $m^{\prime }$, then he must also
visit every $m^{\prime \prime }$ between them. Therefore, if $m$ and $%
m^{\prime }$ are recurrent, so is $m^{\prime \prime }$. Suppose $\overline{m}%
>\mu /c$. Then, there is a profitable deviation for the agent that instructs
him to remain at $\overline{m}-1$ whenever the original strategy instructs
him to switch to $\overline{m}$. $\square $\medskip

\noindent \textbf{Step 3:} \textit{There is a best-reply that induces an
invariant distribution that assigns positive probability to exactly two
values of} $m$.

\noindent \textit{Proof}. Consider the invariant distribution over $\left(
d,m\right) $ induced by the trainer's strategy and the agent's best-reply.
By Step 1, $\overline{m}-\underline{m}\geq 1$. If $\overline{m}-\underline{m}%
=1$, we are done. Therefore, assume $\overline{m}-\underline{m}>1$. There
are two cases to consider.

First, let $\alpha =1$ (this fits the case of $c\geq 1/2$). This means that
whenever $s=L$, the state switches immediately to $s=H$ in the next period.
Consider the top two values of $m$ in the invariant distribution, namely $%
\overline{m}$ and $\overline{m}-1$. By Step 2, $\overline{m}\leq \mu /c$.
Moreover, when $s=L$ (at which $d$ attains its lowest value according to the
trainer's strategy), the agent strictly prefers $\overline{m}-1$ to $%
\overline{m}$. Consider some $t$ for which $m_{t}=\overline{m}$ (there are
infinitely such periods because $\overline{m}$ is recurrent). If $s_{t+1}=L$%
, the agent necessarily switches to $m_{t+1}=\overline{m}-1$. If, on the
other hand, $s_{t+1}=H$, we need to consider two possibilities.\medskip

\begin{itemize}
\item Suppose that when $s_{t+1}=H$, it is not optimal for the agent to play 
$m_{t+1}=\overline{m}$. That is, the agent switches from $m_{t}=\overline{m}$
to $m_{t+1}=\overline{m}-1$ for \textit{any} realization of $s_{t+1}$. But
this also means that if $m_{t^{\prime }}=\overline{m}-1$ at some period $%
t^{\prime }$ and $s_{t^{\prime }+1}=H$, it cannot be optimal for the agent
to switch to $m_{t^{\prime }+1}=\overline{m}$. The reason is that by
revealed preference, the agent prefers being at $\overline{m}-1$ to being at 
$\overline{m}$ when the state is $H$. And since we already saw that the
agent prefers being at $\overline{m}-1$ to being at $\overline{m}$ when the
state is $L$, this means that the agent will \textit{never} switch from $%
\overline{m}-1$ to $\overline{m}$, contradicting the definition of $%
\overline{m}$ as a recurrent state.\medskip

\item Suppose that when $s_{t+1}=H$, it is optimal for the agent to play $%
m_{t+1}=\overline{m}$. This reveals a weak preference for $\overline{m}$
over $\overline{m}-1$ when the state is $H$. Therefore, there is a
best-reply for the agent that prescribes $m_{t+1}=\overline{m}$ whenever the
extended state $(s_{t+1},m_{t})$ is $(H,\overline{m}-1)$ or $(H,\overline{m}%
) $. We already saw that when the extended state is $(L,\overline{m})$, the
agent switches to $\overline{m}-1$. Since $\alpha =1$, this means that we
have constructed a best-reply for the agent such that once he reaches $%
\overline{m}$, he will only visit $\overline{m}$ and $\overline{m}-1$ from
that period on, contradicting the assumption that there are additional
recurrent values of $m$.\medskip
\end{itemize}

Thus, we have ruled out the possibility that $\overline{m}-\underline{m}>1$
when $\alpha =1$. Now suppose $\beta =1$ (this fits the case of $c\leq 1/2$%
). An analogous argument establishes that there is a best-reply for the
agent that induces an invariant distribution with only two recurrent mass
values, $\underline{m}$ and $\underline{m}+1$.

It follows that we can restrict attention to strategies of the agent that
induce an invariant distribution which assigns positive probability to
precisely two consecutive mass values, $m$ and $m-1$, where $0<m\leq \mu /c$%
. $\square $\medskip

\noindent \textbf{Step 4:} \textit{There is a best-reply for the agent that
induces an invariant distribution on the mass values }$\mu /c$\textit{\ and }%
$\mu /c-1.$

\noindent \textit{Proof. }Given Step 3, it is clearly optimal for the agent
to be at $m$ when $s=H$ and at $m-1$ when $s=L$. In addition, when $m>\mu /c$
($m<\mu /c-1$), the agent clearly wants to move downward (upward).

The invariant distribution of the trainer's two-state Markov process assigns
probability $\alpha /(\alpha +\beta )$ to state $H$ and $\beta /(\alpha
+\beta )$ to state $L$. Therefore, since the agent is arbitrarily patient,
his long-run expected payoff is approximately%
\[
-\frac{\alpha }{\alpha +\beta }\cdot (cm+\frac{\mu }{c}-m)-\frac{\beta }{%
\alpha +\beta }\cdot c(m-1) 
\]%
It is now easy to see that given that $\alpha /(\alpha +\beta )>c$, this
expression increases with $m$, such that the optimal value of $m$ is $\mu /c$%
. The expected value of $m$ according to this strategy is%
\[
\frac{\alpha }{\alpha +\beta }\cdot \frac{\mu }{c}+\frac{\beta }{\alpha
+\beta }\cdot (\frac{\mu }{c}-1) 
\]%
which is arbitrarily close to the upper bound. $\blacksquare $

\section{Discussion}

In this section we discuss two features of our model.

\subsection{The Importance of Randomization}

Randomization is a feature of the optimal training strategy in our model. In
the myopic-agent case, it ensures a unique invariant mass distribution.
Randomization plays a different role in the patient-agent case. In
particular, when $c<\frac{1}{2}$, a rest period is followed by another rest
period with positive probability. Is this randomization necessary? Or can
the same long-run mass be sustained by a deterministic strategy with the
same long-run distribution over $d$? The following example illustrates that
the answer is negative.

Suppose $\mu =4$ while $c$ is slightly below $\frac{4}{11}$. Then, the
optimal training strategy we presented in Proposition \ref{prop2} induces an
invariant distribution that assigns probability $\frac{4}{11}$ to $d=11$ and
probability $\frac{7}{11}$ to $d=0$. The strategy sustains a minimal
long-run mass level of $m=10$.

Now consider a deterministic strategy that induces the same long-run
frequencies of $d$. The strategy follows an $11$-period cycle consisting of
four consecutive periods of $d=11$ and seven consecutive periods of $d=0$.
If the agent plays $m=11$ when $d=11$ and $m=10$ when $d=0$ - as he does
against the strategy presented in Proposition \ref{prop2} - the minimal
long-run mass is $m=10$. However, given the predictable evolution of $d$
under the cyclic deterministic strategy, an agent with $\delta \rightarrow 1$
can do better. Suppose that he plays the following sequence of $m$ against
the cyclic sequence of $d$:%
\[
\begin{array}{cccccccccccc}
d & 11 & 11 & 11 & 11 & 0 & 0 & 0 & 0 & 0 & 0 & 0 \\ 
m & 11 & 11 & 11 & 10 & 9 & 8 & 7 & 7 & 8 & 9 & 10%
\end{array}%
\]%
Compared with the benchmark strategy of playing $m=11$ ($10$) against $d=11$
($0$), the agent saves approximately%
\[
c\cdot (1+1+2+3+3+2+1)-1\approx \frac{41}{11}
\]%
per cycle. Even if this is not a best-reply to the cyclic deterministic
strategy, it means that any best-reply will lead to a minimal long-run mass
below $m=10$.

This example highlights a key role of the stochasticity of the trainer's
optimal strategy. The fact that there is always a chance that the agent will
be required to exert high effort following a rest period incentivizes the
agent not to lower his mass. In contrast, the predictable nature of the
cyclic deterministic strategy allows the agent to gradually lower his mass
and gain it back by the time he needs to exert effort. In particular, it is
profitable for the agent to lower his mass already in the final period of
the high-intensity phase of the cycle, even though this involves a costly
performance gap, because this is more than offset by the cumulative
maintenance-cost saving over the cycle. This is reminiscent of the
phenomenon known as \textquotedblleft overtraining\textquotedblright , where
an individual's performance begins to deteriorate during the high-intensity
phase of a periodization strategy (see Cadegiani and Kater (2019)). The
optimal stochastic strategy avoids this effect.

\subsection{The Trainer's max-min Criterion}

In our model, the trainer's objective is to maximize the agent's minimal
long-run mass. Alternatively, we could use the long-run $average$ mass as a
criterion. However, this criterion is less attractive in our context because
it does not reflect the idea of \textquotedblleft
preparedness\textquotedblright\ - namely, that the body should be able to
perform at a \textit{consistently} high level. In particular, the average
criterion allows zero to be a recurrent value for the agent's mass (and
consequently, his level of preparedness).

A by-product of our analysis in Section 3 is that in the myopic-agent case, $%
2\mu $ is an upper bound on the average long-run mass that the trainer can
attain. It can be shown that this upper bound can be approximated
arbitrarily well, but this must come at the price of arbitrarily long
recurrent stretches of $m_{t}=0$ realizations (which are compensated for by
periods in which $m_{t}$ reaches arbitrarily high values). Obviously, such
paths imply that the agent cannot consistently meet positive challenge
levels. By comparison, the process we constructed in Section 3 induces an
average long-run mass of approximately $2\mu -\frac{1}{2}$ and a minimal
long-run mass of $2\mu -1$.

A similar diagnosis pertains to the patient-agent case (we treat $\mu $ as a
precise upper bound on average intensity, for the sake of the argument). An
upper bound on the average long-run mass is $\mu /c$. The reason is that if
average mass exceeds this value, it implies that the agent's average
long-run cost is above $\mu $. However, the agent can ensure an average cost
of $\mu $ by always playing $m=0$, hence a long-run mass in excess of $\mu /c
$ is inconsistent with the agent's best-replying. We believe that as in the
myopic-agent case, this upper bound can be approximated arbitrarily well.
However, as in the myopic-agent case, recurrent stretches of $m_{t}=0$
realizations are necessary for this - which, once again, fails the max-min
criterion miserably. By comparison, the process we constructed in Section 4
induces an average long-run mass of approximately $\mu /c-1+c$, and a
minimal long-run mass of $\mu /c-1$.

\section{Conclusion}

In this paper we presented a theoretical approach to the subject of exercise
physiology, based on the view of the human body as a forward-looking
optimizing agent which is nevertheless constrained by sluggish adjustment.
We saw that this very sluggishness is actually a boon to physical trainers:
using a stochastic training strategy that resembles popular
\textquotedblleft periodization\textquotedblright\ techniques, the trainer
can achieve a significantly higher long-run muscle mass than if the body
could instantaneously adjust its mass to physical stress.

Our analysis focused on the two polar cases of $\delta =0$ and $\delta
\rightarrow 1$. While the optimal strategy is similar in the two cases, the
logic that sustains them is different. Therefore, finding the optimal
strategy for arbitrary $\delta \in (0,1)$ remains an open problem.

We believe that thanks to its abstraction, our modeling approach can be
extended to related problems, such as the optimal design of dynamic dieting
regimes. A model that describes the body's metabolism as a consequence of
dynamic $sluggish$ optimization with rational expectations may shed light on
prevalent dieting programs such as carb cycles. We hope to pursue this
approach before the next pandemic.

\end{document}